\documentclass[11pt,twoside]{article}
\usepackage{graphicx}
\usepackage{ihep,epsf,amsfonts,amssymb,amsbsy}

\begin{document}
\title
%\begin{center}
{\Large\bf \uppercase{Ghost condensate model \\ of flat rotation curves}} %\\[4mm]
\author{
{\bf V.V.Kiselev }\\ [1mm] %\email{kiselev@th1.ihep.su}
{\it \small Russian State Research Center ``Institute for High Energy
Physics'', Protvino, Russia}
}
\date{}
\maketitle

%\end{center}

%\pacs{95.35.+d,  98.62.Gq,
%04.40.-b, 04.40.Nr}
%\maketitle

\begin{abstract}
An effective action of ghost condensate with higher derivatives
creates a source of gravity and mimics a dark matter in spiral
galaxies. We present a spherically symmetric static solution of
Einstein--Hilbert equations with the ghost condensate at large
distances, where flat rotation curves are reproduced in leading
order over small ratio of two energy scales characterizing
constant temporal and spatial derivatives of ghost field:
$\mu_*^2$ and $\mu_\star^2$, respectively, with a hierarchy
$\mu_\star\ll \mu_*$. We assume that a mechanism of hierarchy is
provided by a global monopole in the center of galaxy. An estimate
based on the solution and observed velocities of rotations in the
asymptotic region of flatness, gives $\mu_*\sim 10^{19}$~GeV and
the monopole scale in a GUT range $\mu_\star\sim 10^{16}$~GeV,
while a velocity of rotation $v_0$ is determined by the ratio: $
\sqrt{2}\, v_0^2= \mu_\star^2/\mu_*^2$. A critical acceleration is
introduced and naturally evaluated of the order of Hubble rate,
that represents the Milgrom's acceleration.
\end{abstract}

\vskip 5mm
\hbadness=1500
%\end{document}
%%\section{Introduction}
%
In addition to cosmological indirect indications of dark matter
representing a nonbaryonic pressureless contribution to energy
budget of Universe during evolution \cite{WMAP,SNe}, there are
explicit observational evidences in favor of existence of dark
matter. Firstly, rotation curves in spiral galaxies cannot be
explained by Keplerian laws with visible distributions of luminous
baryonic matter at large distances, where curves are becoming flat
and reveal a $1/r^2$-profile of mass for the dark matter at large
distances, if Newtonian dynamics remains valid. Secondly,
gravitational lensing by galaxies corresponds to masses, which are
significantly greater than those of visible matter. Thirdly,
virial masses in clusters of galaxies witness for the dark matter,
too. While the existence of dark matter is well established, its
nature and origin are under question \cite{dark,silk}.

The most straightforward opportunity is to assume an existence of
weakly interacting massive particle, which could be experimentally
observed in on-Earth-grounded facilities \cite{silk}. However,
numerical simulations of N-body dynamics for the cold dark matter
\cite{NFW} give, firstly, a more rapid decay of mass density with
distance ($1/r^3$ instead of $1/r^2$) and, secondly, a $1/r$-cusp
in centers of galaxies. Yet both phenomena are in contradiction
with observations, which prefer a core-like distribution with a
constant density of dark matter in the center \cite{Salucci} and
do not exhibit a falling down of rotation curves in spiral
galaxies.

A second way suggests a modification of Newtonian dynamics (MOND)
in the asymptotic regions of flat rotation curves. The most
successful approach was offered by Milgrom in his MOND
\cite{Milgrom,BM,Sanders}, which has many phenomenological
advances. Milgrom supposed a phenomenon of critical acceleration
$a_0$ below which the gravitational dynamics should be modified in
order to reproduce the flat rotation curves, so that an actual
acceleration is equal to $\sqrt{g_N a_0}$, where $g_N$ is the
acceleration generated by visible matter in galaxy. In the
framework of MOND the rotation curves can be explained in terms of
visible matter only! Moreover, the critical acceleration naturally
leads to a strong correlation of asymptotic velocities with
visible masses of galaxies: the Tully--Fisher law. Similar
successes of MOND are reviewed in \cite{SanMcG}. Certainly, the
phenomenological evidence in favor of critical acceleration
challenges the model of cold dark matter, where a regularity like
the Tully--Fisher law seems quite occasional and could be some-how
introduced as an effect of evolution only. Some theoretical
shortcomings of primary MOND model \footnote{For instance, one
could mention superluminal velocities of graviscalar and
insufficient gravitational lensing.} have been recently removed in
a novelty version of tensor-vector-scalar theory by Bekenstein
\cite{Bekenstein}. However, a critical acceleration remains an
\textit{ad hoc} quantity in MOND paradigm as an indication of
essential modification of general relativity in infrared.

Another example of modification is a nonsymmetric gravitation
theory by Moffat \cite{Moffat}, which involves several parameters,
for example, a decay length of extraordinary force. The parameters
can be also combined to compose a critical acceleration. A
question is whether an introduction of nonsymmetric rank-2 tensor
instead of metric is natural or rather exotic.

Finally, a possible explanation of flat rotation curves in terms
of scalar fields was tried in several papers
\cite{Arbey,NSS,Matos}. One found that relevant scalar fields
should differ from both a scalar quintessence~\cite{quint}
responsible for a measured acceleration of Universe expansion
\cite{SNe} and a scalar inflaton governing an inflation in early
Universe \cite{introduc}. For example, Nucamendi, Salgado and
Sudarsky (NSS) \cite{NSS} have derived a metric consistent with
flat rotation curves caused by a presence of perfect fluid given
by a scalar field. Moreover, they have found that the scalar field
should be represented by a triplet with an asymptotic behavior of
global monopole at large distances. In addition, the NSS metric is
consistent with gravitational lensing, too. Nevertheless, one
still has not found a convincing relation of parameters in a
scalar field dynamics with properties of rotation curves.

Let us focus an attention on the rotation curves. In present
letter we introduce a ghost condensate model which dynamical
parameters are deeply related with characteristics of rotation
curves. Moreover, we find a natural way to get a critical
acceleration in general relativity with the ghost condensate and
estimate its value, which turns out to be of the order of Hubble
rate at present day in agreement with phenomenological
measurements.

The ghost condensate \cite{ACLM} is an analogue of Higgs
mechanism. Indeed, a tachyon field $\sigma$ with a negative square
of mass can be stabilized by $\lambda \sigma^4$ term of its
potential, which leads to a tachyon condensate, known as a Higgs
mechanism in gauge theories. Similarly, a ghost field $\phi$
possessing an opposite sign of kinetic term can be stabilized by
introduction of higher order terms leading to a ghost condensate.
In contrast to tachyon condensation being a renormalizable and
Lorentz-invariant procedure, an isotropic homogeneous ghost
condensation gives a nonzero square of time derivative $\langle
\dot\phi^2\rangle$, which breaks a Lorentz invariance, while
higher derivative terms are acceptable in an effective theory in
infrared, only. As for the breaking down the Lorentz invariance,
it can simply imply appearing an arrow of time in a non-static
isotropic homogeneous expanding Universe with ordinary
Friedmann--Robertson--Walker metric. A modification of gravity in
infrared by postulating a Goldstone nature of ghost in an
effective theory was investigated in \cite{ACLM}. This model leads
to instability of gravitational potential in a time exceeding the
Universe age at least \cite{Dubovsky,Peloso}. We do not accept the
Goldstone hypothesis, that allows us to avoid strict constraints
on dimensional parameters of ghost action. A dilatonic ghost
condensate as dark energy is considered in \cite{add}.

A leading term of lagrangian for the ghost field with invariance
under a global translations $\phi\to \phi+ c$ is given by
\begin{equation}\label{P_X}
    {\cal L}_X = P(X), \qquad X=\partial_\nu\phi\partial^\nu\phi,
\end{equation}
in flat space-time with a metric signature $(+,-,-,-)$, so that $
{\cal L}_X\to -\frac{1}{2}\,\partial_\nu\phi\partial^\nu\phi$ at $
X\to 0$ reproduces the kinetic term with the negative sign,
indicating instability which will be removed by ghost
condensation. Indeed, expanding (\ref{P_X}) near $X_0 =\mu_0^4$ by
fixing
$$
\phi =\mu_0^2\, t+\pi(x),\quad \partial_\nu\phi =(\mu_0^2+\dot
\pi,\boldsymbol \nabla \pi),
$$
we get a quadratic approximation for small perturbations %%$\pi$
$$
{\cal L}_X^{(2)} =\dot\pi^2
(P^\prime+2\mu_0^4\,P^{\prime\prime})-(\boldsymbol \nabla
\pi)^2\,P^\prime,
$$
which is stable at $P^\prime>0$ and $P^\prime+2\mu_0^4\,
P^{\prime\prime}>0$ at any suitable $X_0$. For definiteness, at
relevant values of $X$ we put a Higgs-like function
\begin{equation}\label{canon}
P(X) = -\frac{m_0^2}{2\eta^2}\, X+\frac{\lambda}{4\eta^4}\,X^2.
\end{equation}
An expansion with the Friedmann-Robertson--Walker metric gives
$$
\partial_t\left(a(t)^3\,P^\prime\,\dot\phi
\right)=0\quad\Rightarrow\quad P^\prime\,\dot\phi=\frac{{\rm
const.}}{a^3}\to 0,
$$
where $a(t)$ is a scale parameter of metric. So, the evolution
drives to $P^\prime\to 0$, since $\dot\phi=0$ is not a stable
point by construction. Thus, a preferable choice is an extremum
point $P^\prime(X_0) = 0$ with $P^{\prime\prime}> 0 $ \footnote{It
is easy to recognize that a substitution of $\partial_\nu\phi
=\eta\, {\cal A}_\nu$ transforms lagrangian (\ref{P_X}) to a
potential of vector field ${\cal A}_\nu$, and the preference point
is the extremum of potential for the vector field, representing
the ghost condensate (see also \cite{me}).}. We introduce a
correction of the form
\begin{equation}\label{corr}
    \Delta {\cal L}
    =-\frac{1}{2\eta^2}\,\partial_\alpha\partial_\beta\phi\,
    \partial^\alpha\partial^\beta\phi,
\end{equation}
which does not destroy a stability, since in quadratic
approximation it gives
$
\Delta{\cal L}^{(2)} \approx -\frac{1}{2\eta^2}\,(\boldsymbol
\nabla^2\pi)^2,
$
leading to the following dispersion relation for perturbations
$\pi$ in momentum space: $\omega^2\approx\boldsymbol k^4/2m_0^2.$
A scaling analysis performed in \cite{ACLM} has confirmed that the
model is a correct effective theory in infrared.

Next, consider the ghost condensate in presence of global monopole
\cite{monopole}. Then we put \textit{at large distances}
\begin{equation}\label{ghost}
    \phi=\mu_*^2\,t-\mu_\star^2\,r+\sigma(x),
\end{equation}
with $ \mu_*^2-\mu_\star^2=\mu_0^2,$ $
\kappa={\mu_\star}/{\mu_*}\ll 1, $ so $P^\prime =0$, and we add a
correction induced by monopole $ \Delta\widetilde{\cal L}
=-\varkappa^2\left(\boldsymbol \nabla\phi+\boldsymbol
n\,\mu_\star\right)^2,\, \boldsymbol n=\boldsymbol \nabla r, $
where $\mu_\star$ fixes an energy scale in dynamics of monopole,
while $\varkappa^2>0$ guarantees a stability of monopole, and its
rather large absolute value preserves a stability over
perturbations \footnote{In Minkowski space-time at large distances
from a monopole center, one could compose a constant four vector
${\cal A}^\nu$ by the ghost field and monopole triplet-scalar
\cite{Vilenkin}, so that the temporal derivative of ghost
$\dot\phi$ would be combined with the spatial triplet
$\phi^a=\boldsymbol n$ in $ {\cal A}^\nu
=\frac{1}{\eta}\,\{\dot\phi,\mu_\star^2\boldsymbol n\}, $ which
take the form $ {\cal A}^\nu
=\frac{1}{\eta}\,\{\mu_*^2,\mu_\star^2,0,0\} $ in polar
coordinates $\{t,r,\theta,\phi\}$ with the ghost
$\phi=\mu_*^2\,t$. In Minkowski space-time we can simply put $
\eta\,{\cal A}_\nu =\partial_\nu(\mu_*^2\, t -\mu_\star^2\,r), $
which is exact in this case, and we get a purely gauge vector
field \textit{composed by the ghost in presence of global
monopole} (see (\ref{ghost})).}, \footnote{Higgs-like fields as
dark matter are treated in \cite{Bert}.}.

Neglecting perturbations, we study the ghost condensate in
presence of monopole (\ref{ghost}) as a source of gravity at large
distances in spherically symmetric quasi-static limit.
%\footnote{The static limit is an approximation based on a
%factorization of scales: the time scale in evolution of Universe,
%given by inverse Hubble rate, is much greater than sizes of
%galaxies. The approximation is tested below.}.
Then, the only source of energy-momentum tensor is the correction
of (\ref{corr}), where we should replace partial derivatives by
covariant ones \footnote[1]{A model extension to a curved
space-time is the following: $
%\begin{equation}\label{curved}
    {\cal A}^\nu
    =\frac{1}{\eta}\,\{\mu_*^2,\mu_\star^2,0,0\},\quad
    \Delta{\cal L} =-\frac{1}{2}\,\nabla_\alpha{\cal A}^\nu\nabla^\alpha{\cal
A}_\nu,
$ %\end{equation}
i.e. the constant covariant four-vector is reasonably motivated,
%Note, that with a constant spatial components of curved metric
%further obtained, the triplet-scalar contribution is recovered
%from (\ref{curved}) up to a normalization factor $1+{\cal
%O}(v_0^2)$,
though the specified ${\cal A}^\nu$ cannot be
represented as a gradient function, since $ {\cal A}_{t;r}-{\cal
A}_{r;t}\neq 0. $} with the metric
\begin{equation}\label{static}
    {\rm d}s^2 ={\tt f}(r)\,{\rm d}t^2-\frac{1}
    {{\tt h}(r)}\,{\rm d}r^2-r^2[{\rm
    d}\theta^2+\sin^2\theta\,{\rm d}\varphi^2],
\end{equation}
so that due to a small parameter $\kappa$ we can expand in it and
find the following solution of corresponding Einstein--Hilbert
equations in the leading order over $\kappa$ \footnote{We use an
ordinary notation for the derivative with respect to the distance
by the prime symbol $\partial_r f(r) =f^\prime(r)$.}:
\begin{equation}
    v_0^2=\kappa^2/\sqrt{2},\quad {\tt h}(r) =
    1-2v_0^2,\quad{\tt f}'(r)=\frac{2v_0^2}{r},
\end{equation}
giving a $1/r^2$-profile of the curvature and a flat asymptotic
behavior for a constant velocity of rotation $v_0$ \footnote{Thus,
we have found the NSS metric with a perfect fluid of ghost
condensate.}, if
$
    8\pi\,G\, %%%\frac
    {\mu_*^4}/{\eta^2}=1+{\cal O}(\kappa^4).
$
In the leading order at $\mu_\star\ll \mu_*$ we have
$\mu_*^4\approx m_0^2\eta^2/\lambda$, and, hence, $8\pi\,G\,
m_0^2/\lambda\approx 1$. So, putting $\eta=m_0$ for a canonical
normalization in (\ref{canon}) at $\lambda\sim 1$, we get
\newpage

\begin{equation}\label{constar}
    8\pi\,G\,\mu_*^2\sim 1,
\end{equation}
and the ghost condensate scale is of the order of Planck mass:
$\mu_*\sim 10^{18-19}$ GeV.

The solution leads to the temporal component of the
energy-momentum tensor dominates and has the required profile with
the distance \footnote{The accretion of ghost to the center of
gravity should be suppressed at $P'=0$ (see \cite{AFrolov}).}:
\begin{equation}
 T_r^r\sim T_\varphi^\varphi\sim T_\theta^\theta\sim {\cal
 O}(v_0^2)\cdot T_t^t\sim {\cal O}\left(\frac{1}{r^2}\right).
\end{equation}
Numerically at $v_0\sim 100-200$ km/sec, we get
\begin{equation}
    \kappa\sim 10^{-3}\quad \Rightarrow\quad \mu_\star\sim 10^{15-16}\,\mbox{GeV,}
\end{equation}
so, the characteristic scale in the dynamics of monopole is in the
range of GUT. Thus, the small ratio of two natural energetic
scales determines the rotation velocity in dark galactic halos.

Since we treat the ghost condensate as an external source in the
Einstein--Hilbert equations, let us consider conditions providing
that the corrections could be neglected.

Firstly, suppressing the dependence of ghost on the distance, in
the Friedmann-Robertson--Walker metric we find that the covariant
derivatives in (\ref{corr}) generate the correction, determined by
the Hubble rate $H$ (see \cite{me}):
\begin{equation}
    \delta {\cal L}=-\frac{3}{2\eta^2}\,H^2 \,\dot\phi^2,
\end{equation}
so that the temporal derivative acquires a slow variation with the
time due to the displacement of stable point, since we can use an
effective quantity $m^2_{\rm eff} = m_0^2+3H^2$, and the
dependence is really negligible, if the Hubble rate is much less
than the Planck mass, $H\ll m_0\sim m_{\rm Pl}$.
%Note, that as was
%shown in \cite{me}, the dependence of ghost condensate on time
%generates a variation of effective gravitational constant with the
%time in the evolution equations. However, such the dependence
%should be small as forced by the experimental observations. We
%have seen that in the scheme described above the time-dependence
%of ghost condensate is negligible.

Secondly, if we take into account both the expansion and radial
dependence, then in presence of monopole the covariant derivatives
of ghost with respect to polar coordinates $(r,\theta,\varphi)$
(more accurately see [31])
\begin{equation}
\label{if}
    \phi^{;r}_{\;;r}=H\,\dot\phi,\qquad
    \phi^{;\theta}_{\;;\theta}=\phi^{;\varphi}_{\;;\varphi}=-\frac{1}{r}\,\phi^\prime
    +H\,\dot\phi,
\end{equation}
induce the correction
\begin{equation}
\label{if-V}
    \delta_\star {\cal L}
    =-\frac{1}{2\eta^2}\,H^2\,\dot\phi-\frac{1}{\eta^2}\left(-\frac{1}{r}\,\phi^\prime+H\,\dot\phi\right)^2,
\end{equation}
which can be neglected at large distances, only. Therefore, the
`cosmological limit' of ghost condensate is consistently reached,
if
\begin{equation}
    \frac{1}{r}\,\mu_\star^2\ll H\,\mu_*^2.
\end{equation}
The consideration above is disturbed because of (\ref{if-V}) at
distances less than $r_0$ defined by
\begin{equation}
\label{r0}
    \frac{1}{r_0}\,\mu_\star^2 = \varepsilon\,H\,\mu_*^2\quad \Rightarrow\quad
    \frac{1}{r_0}\,\frac{\mu_\star^2}{\mu_*^2} = \varepsilon\,H_0,
\end{equation}
where $H_0=H(t_0)$ is the value of Hubble rate at present, and
$\varepsilon$ is a parameter of the order of $1-0.1$. Substituting
$\mu_\star^2/\mu_*^2=\sqrt{2}\,v_0^2$ into (\ref{r0}), we get $
%\begin{equation}\label{acc}
    {v_0^2}/{r_0} ={\varepsilon\,}\,H_0/{\sqrt{2}},
$ %\end{equation}
while
\begin{equation}\label{a0}
    a_0 =\frac{v_0^2}{r_0}
\end{equation}
is a centripetal acceleration, and, then, the critical
acceleration is determined by the Hubble rate,
\begin{equation}\label{Milgrom}
    a_0 =\frac{\varepsilon\,}{\sqrt{2}}\,H_0,
\end{equation}
that is the acceleration below which the limit of flat rotation
curves becomes justified. That is exactly a direct analogue of the
critical acceleration introduced by Milgrom in the framework of
MOND \cite{Milgrom}.

Further, we could suppose that in the case, when the gravitational
acceleration produced by the visible matter in the galactic
centers exceeds the critical value, we cannot reach the limit of
flat rotation curves. Indeed, in that case the distance dependence
cannot be excluded from the ghost condensate. The Newtonian
acceleration at distance $r_0$ is equal to
\begin{equation}
    a_0^* =\frac{G {\cal M}}{r_0^2},
\end{equation}
where $\cal M$ is a visible galactic mass. According to
(\ref{Milgrom}), the critical acceleration is a universal quantity
slowly depending on the time, while (\ref{a0}) implies that the
distance and velocity can be adjusted by variation in order to
compose the universal $a_0$. Therefore, we should put
\begin{equation}
\label{coin}
    a_0^* =a_0,
\end{equation}
which yields
\begin{equation}\label{Tully-Fisher}
    v_0^4 = G{\cal M} a_0.
\end{equation}
The galaxy mass is proportional to an H-band luminosity of the
galaxy $L_H$, so that (\ref{Tully-Fisher}) reproduces the
Tully--Fisher law
$ %\begin{equation}
    L_H \propto v_0^4.
$ %\end{equation}
Then, other successes of MOND can be easily
incorporated in the framework under consideration, too.

Nevertheless, we could treat (\ref{coin}) as a coincidence. For
instance, (\ref{Tully-Fisher}) leads to
$$
\mu_\star^4\sim {\cal M} H_0/G.
$$
Therefore, if the flattening is observed in a spiral galaxy, the
mass of galaxy should strongly correlate with the Hubble rate at
present as well as the scale of monopole dynamics. This fact could
be reflected in a correlation of rotation velocity with a mass of
central body, a supermassive black hole, as observed empirically.

Thus, we have presented the working example of ghost condensate
model in presence of monopole in order to get the description of
flat rotation curves in spiral galaxies at large distances. There
are two energy scales in the model. The scales are natural, and
they represent the Planck mass and GUT scale. The critical
acceleration determining the region of validity for the model has
been estimated in general relativity with the ghost condensate.

This work is partially supported by the grant of the president of
Russian Federation for scientific schools NSc-1303.2003.2, and the
Russian Foundation for Basic Research, grant 04-02-17530.

%\newpage
%%\section*{References}

\end{document}